\documentclass[11pt]{article}
\usepackage{paper_c_style}
\title{Causal Effects of Protocol-Fee Changes on Liquidity Provision in Automated Market Makers}
\author{Wen-Ting Wang}
\date{July 2026\\
{\small Email: \texttt{egpivo@gmail.com}}}
\newcommand{\E}{\mathbb{E}}

\begin{document}
\maketitle

\begin{abstract}
Automated market maker (AMM) fee rules are often evaluated by liquidity-provider (LP)
welfare, but that objective mixes fee revenue, adverse-selection loss
(loss-versus-rebalancing, LVR), routing response, and liquidity supply. Fixed-fee
Uniswap~v3 history cannot separate these channels or identify counterfactual trader-facing
dynamic-fee rules. Real fee-related
variation nonetheless exists: the Uniswap protocol-fee switch (executed 2025-12-28) cut LP
take-rates with tier-differentiated intensity while leaving trader-facing fees unchanged.
Using a pre-specified matched-overlap event-study difference-in-differences (DiD) design, we
estimate the liquidity-supply response to take-rate cuts, the kernel $K_L$ that
simulator-based fee-controller evaluations routinely freeze. Treatment, event time, unit
roles, and outcomes are reconstructed from public logs into a frozen, hash-checked panel
before any estimate. We detect no large short-run average response in active liquidity or
local depth; LP participation and composition, more precisely estimated, likewise show
none, so the result is a non-detection at the design's resolution rather than a precise
zero. Token-1 volume and native fee income fail the parallel-trends gate and are reported
descriptively. A channel-admissibility audit delimits the estimand: the LP-side response
$K_L$ is design-based, while trader-facing dynamic-fee protection is a model-conditioned
boundary, not a second estimand.
\end{abstract}


\section{Introduction}\label{sec:intro}

A protocol fee changes the share of trading fees that liquidity providers keep. In an
automated market maker (AMM), this is not only an accounting change: if liquidity
providers (LPs) respond by withdrawing or reallocating capital, the same change also
moves depth, slippage, volume, and fee income. The empirical question is direct: what is
the causal effect of a protocol-fee change on liquidity provision? This is not the
question most fee-controller evaluations answer; they usually compare candidate rules by
expected LP welfare in a specified environment
\citep{campbell2025optimal, baggiani2025optimal, bayraktar2024dexspecs}.

The question matters for decentralized finance (DeFi) market design because AMMs are a
core trading venue and fee policy is set through public protocol rules rather than a
private exchange schedule. A protocol-fee switch reallocates fee income among traders,
LPs, and the protocol treasury while leaving the market-clearing mechanism otherwise
unchanged. Public logs make the resulting trades and liquidity updates observable, but
they do not reveal counterfactual routing, trader types, or LP opportunity costs. The
empirical design must therefore distinguish the margin moved by the policy from the
margins that public data cannot identify.

LP welfare is a mixture. In any accounting of LP outcomes,
\begin{equation}
W \;=\; \mathrm{FeeRev}^{\mathrm{arb}} + \mathrm{FeeRev}^{\mathrm{fund}} - \mathrm{LVR},
\label{eq:welfare}
\end{equation}
where LVR denotes loss-versus-rebalancing. Routing retention and liquidity response act on
these terms from outside. An observed improvement in $W$ can come from reduced
adverse selection, from higher fees on arbitrage, from higher fees on ordinary flow, or
from a response margin the evaluation holds frozen. These are economically different
outcomes with
different design implications, and the objective value does not distinguish them.
Fixed-fee Uniswap~v3 history cannot separate them either: within a pool the trader-facing
fee never moves, cross-tier contrasts confound the fee with pool fundamentals and router
behavior, and the trader types and routing decision sets that define the channels are
latent in public data (Section~\ref{sec:data}).

The Uniswap protocol-fee switch supplies a different source of variation. Executed
2025-12-28, it carved a protocol fee out of LP fee income on a large set of pools, with
tier-differentiated intensity, while leaving trader-facing swap fees unchanged. It
therefore moves the LP take-rate, not the execution price paid by traders, and identifies
an LP-side question: how liquidity provision responds when the LP share of collected fees
is cut (Section~\ref{sec:setting}). It does not identify the trader-facing dynamic-fee
question, because the posted fee tier stays fixed and the required trader and routing
variables are not observed.

The paper estimates this LP-side response, $K_L$, defined as the finite-horizon response path of
active liquidity, depth, and related LP-supply outcomes for treated pools relative to
matched low-exposure controls, with a matched-overlap event-study design, pre-specified
and read causally only under the conditions of Section~\ref{sec:assumptions}. Mechanism
outcomes (slippage, volume, LP composition, just-in-time (JIT) activity, reallocation) are
reported as reduced-form paths, not controls or mediation. Treatment, event time, unit
roles, and outcomes are reconstructed from public logs into a frozen, hash-checked panel
before any estimate: on-chain, the design is only as credible as the deterministic
construction of the objects entering it. The headline finding is a non-detection: active
liquidity and local depth show no large short-run average response, and the more precisely
estimated participation and composition outcomes show none at the design's resolution,
not a precise zero.

This paper makes three contributions. First, it estimates the LP-side liquidity-supply
response to the Uniswap protocol-fee switch with a matched-overlap event-study design
(Sections~\ref{sec:method}, \ref{sec:results}). Second, it reconstructs treatment, event
time, unit roles, and outcomes from public logs into a frozen, hash-checked panel. Third, it
uses the channel map to separate the identified LP-side response from trader-facing
dynamic-fee claims that this shock cannot identify.

\section{Institutional setting and data}\label{sec:instdata}

This section fixes the institutional shock, reconstructs treatment and event time from
on-chain events, and describes the outcome panel and its observability limits;
reconstruction and implementation detail is deferred to Appendix~\ref{app:impl}.

\subsection{The protocol-fee switch}\label{sec:setting}

The UNIfication governance package was proposed in November 2025, passed on 2025-12-25
(125.3M votes for, 742 against), and executed after a two-day timelock on 2025-12-28.
It activated protocol fees carved out of LP fee income: v2 pools moved from 0.30\% all
to LPs to a 0.25\%/0.05\% LP/protocol split; treated v3 pools pay a protocol fee of
$1/4$ of LP fees on the 1 and 5 basis-point (bp) tiers and $1/6$ on the 30\,bp and 100\,bp
tiers. Trader-facing swap fees are unchanged (the shock moves LP revenue share, not
execution prices), which makes it an LP-side take-rate shock targeting the $K_L$
margin: any primary liquidity response operates through LP income rather than through
trader execution prices, with JIT response, LP composition, and cross-venue
reallocation treated as measured mechanism or spillover margins, not ruled out by
assumption. The proposal was public from November 2025; all designs treat the
proposal-to-execution window as an anticipation segment, and the pre-period ends at the
proposal date. The switch is a real policy shock, not a randomized experiment: the
quasi-experimental case rests on the package's protocol-wide, token-politics-driven
timing and on the coverage rule (top fee-generating pools) being selection on
observables at the list margin, licensed by the matched-overlap identification
assumptions of Section~\ref{sec:assumptions}.

\subsection{Treatment, sample, and outcomes}\label{sec:treatmentcon}\label{sec:data}

\begin{figure}[t]\centering
\begin{subfigure}[t]{0.48\textwidth}\centering
\includegraphics[width=\textwidth]{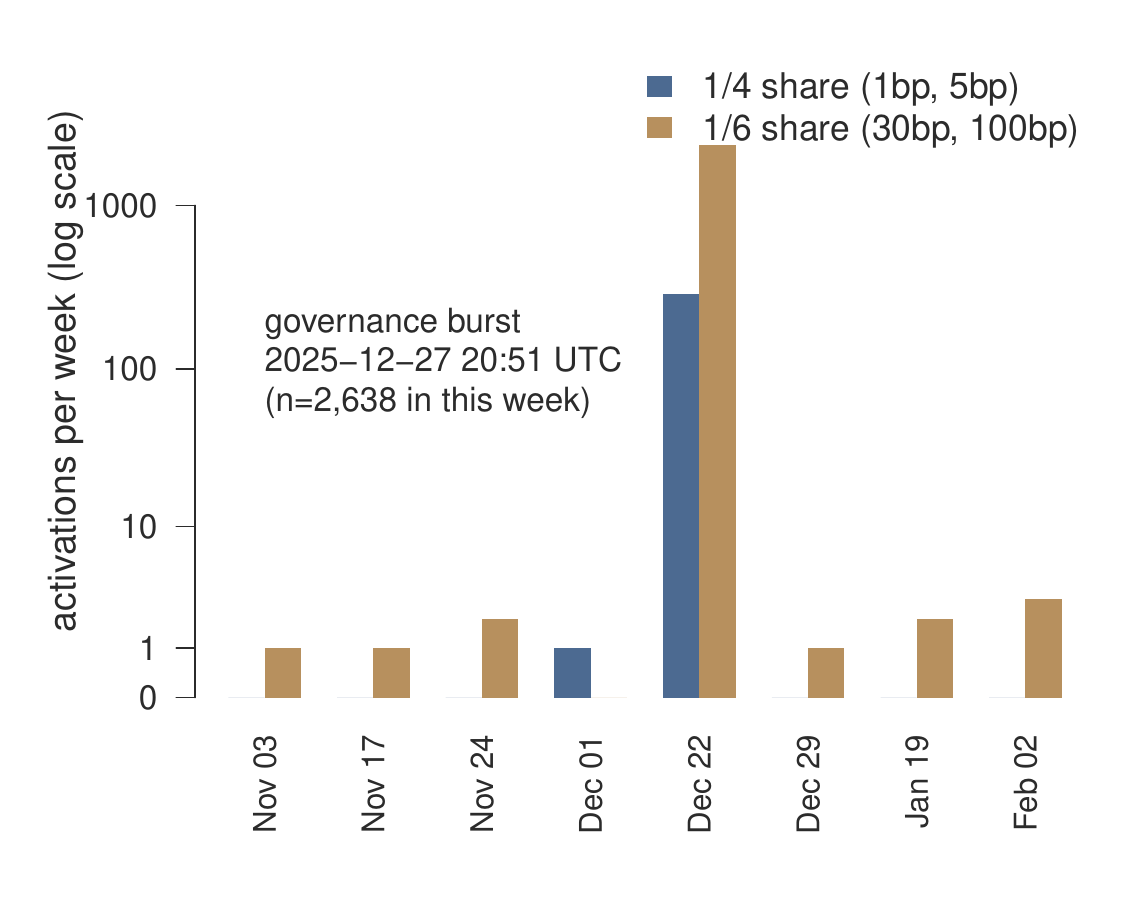}
\subcaption{Weekly activations by intensity group.}\label{fig:activation-a}
\end{subfigure}\hfill
\begin{subfigure}[t]{0.48\textwidth}\centering
\includegraphics[width=\textwidth]{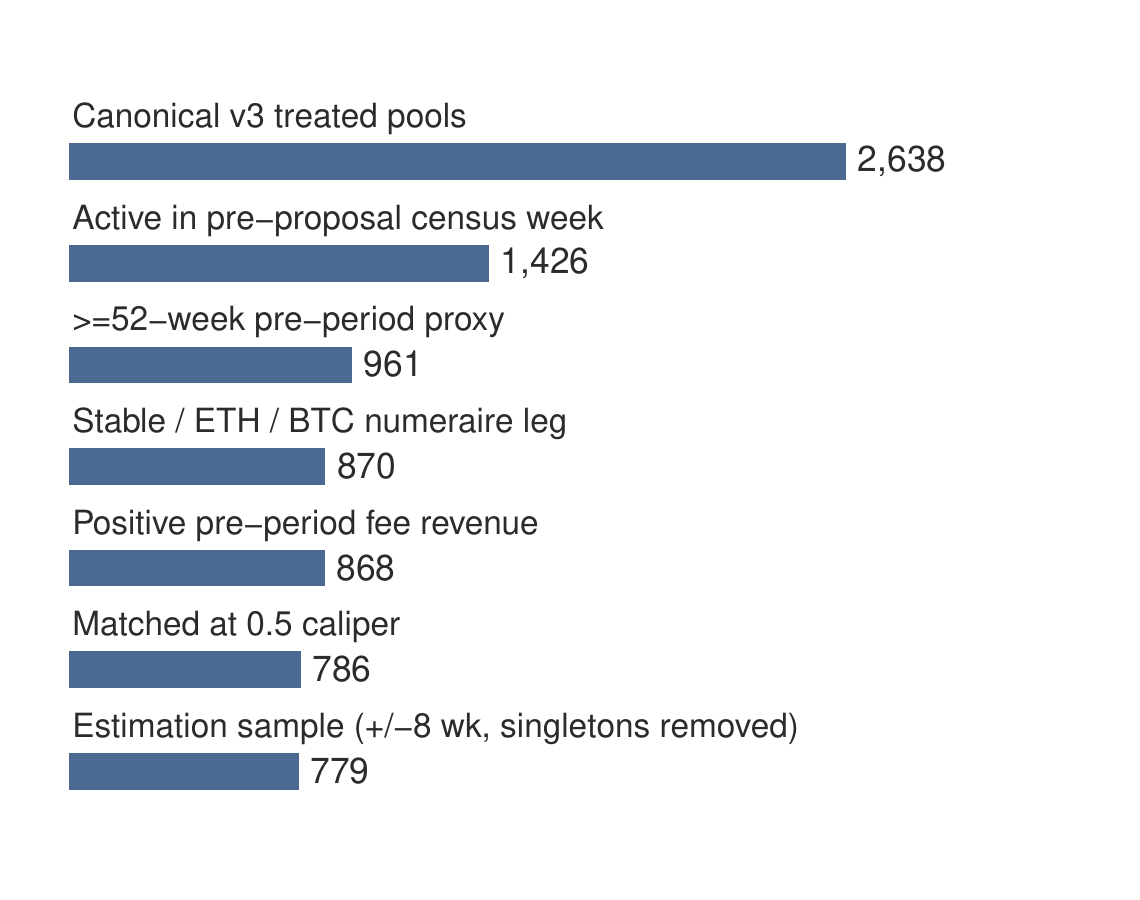}
\subcaption{Design funnel, treated to estimation.}\label{fig:activation-b}
\end{subfigure}
\caption{Treatment activation and sample funnel. (a) Weekly \texttt{SetFeeProtocol}
activations by protocol-share intensity group (1/4: 1\,bp and 5\,bp tiers; 1/6: 30\,bp and
100\,bp tiers), reconstructed from on-chain events on a log count axis; the governance
burst of 2025-12-27 dwarfs the stragglers. (b) The frozen design funnel from 2{,}638
canonical v3 treated pools to the 779-pool estimation sample. Disposition: data
provenance / treatment reconstruction.}
\label{fig:activation}
\end{figure}

The activation inventory can be reconstructed directly from on-chain
\texttt{SetFeeProtocol} events. The governance package is dated 2025-12-28 by
governance-execution convention; on chain, the \texttt{SetFeeProtocol} activation
burst appears from 2025-12-27 20:51 UTC onward (2{,}638 v3 pools in a single batched
burst). Throughout the empirical design, event time $t_0$ is defined by the pool-specific
on-chain activation timestamp, not the governance convention date. Within the burst the
documented intensity rule holds without exception: all 95
one-bp and 192 five-bp pools were set to a $1/4$ protocol share, and all 877 thirty-bp
and 1{,}474 hundred-bp pools to $1/6$. Fork pools emitting the same event signature are
excluded by the canonical-factory filter. A one-week pre-proposal census confirms that some
high-volume untreated pools (e.g.\ 1\,bp correlated-pair pools) stay in the reservoir because
inclusion follows fee-revenue coverage rather than raw volume; the selection variable is
therefore observable, motivating list-margin matching on pre-period realized fee revenue
(Figure~\ref{fig:activation}, Panel A). Treatment objects are read directly from these
events: the indicator $D_i$ from a
canonical pool's activation event, the intensity $\kappa_i \in \{1/4, 1/6\}$ from the
emitted protocol-fee denominator, and the event-time origin $t_{0i}$ from the event's
block timestamp.

The frozen outcome panel is constructed from normalized Swap, Mint, Burn, and Collect
logs, with pool state reconstructed from ordered events; it spans the activation
inventory's treated and control pools and at least one competitor venue over
2024-01--2026-06. The outcome family is fixed before any estimation. The primary outcomes are active
liquidity (time-weighted liquidity within $\pm 1\%$, $\pm 2\%$, and $\pm 5\%$ bands around
mid, $\pm 2\%$ primary) and reconstructed local depth on the same bands. Secondary LP-side
outcomes measure participation and composition: LP entry, exit, unique-LP count, position
duration, same-block JIT share, net liquidity, and collected amount. Flow and revenue
outcomes, namely token-level volume and native LP fee income, are used only to test the
boundary of the causal design. These post-treatment outcomes are never main-specification
controls: conditioning on them in Eq.~\eqref{eq:es} would absorb the response the design
measures. Weeks in which a pool is inactive enter level outcomes as zeros; ratio and
log-type outcomes use the inverse hyperbolic sine, fixed per outcome before estimation.

The activation inventory contains 2{,}638 canonical v3 treated pools; the main
matched-overlap sample is smaller by pre-specified eligibility filters, applied before
any post-period data. Of the 2{,}638, 1{,}426 are active in the pre-proposal census
week; 961 of those were also active a year earlier (the $\geq 52$-week pre-period proxy);
870 carry a stablecoin, ETH, or BTC numeraire leg so that fee revenue in USD is computable;
and 868 have strictly positive pre-period realized fee revenue. These 868 are the treated
main sample; excluded pools carry no usable pre-period panel and are reported in the funnel
(Figure~\ref{fig:activation}, Panel B). Of the 868, 786 (90.6\%) match a canonical-factory
pure control within the 0.5 log-point caliper on pre-period realized fee revenue; the
reservoir holds 622 candidates, 314 used; 82 treated pools are unmatched and reported,
routed to the intensity design and a descriptive treated panel. The design relies on the
dynamic parallel-trends gate, not on level balance; match rates, cell counts, and
standardized mean differences are in Appendix~\ref{app:smd}.

\subsection{Observability limits}\label{sec:margins}

The primary $K_L$ outcomes do not require any trader-side decomposition, and this is what
the public record supports: trader types are latent (swap counterparties are addresses),
router decision sets are not observed (receipts show landed routes only), and the
arbitrage/fundamental attribution of fee revenue is therefore not a measurement. None of
this binds the Block-A liquidity and depth outcomes, but all of it binds any attempt to read
trader-side effects, so trader-facing dynamic-fee effects lie outside the empirical estimand.

\section{Empirical strategy}\label{sec:framework}

This section defines the LP-side causal estimand, the matched-overlap event-study
estimator, and the assumptions under which the event path is read causally. The estimate
is design-based; the channel audit enters only to assign observability and support labels,
explaining why the LP take-rate response is estimable and why trader-facing dynamic-fee
claims remain outside the empirical estimand.

\subsection{Estimand}\label{sec:paths}\label{sec:problem}\label{sec:estimands}

Throughout, $i$ indexes a canonical, factory-filtered pool, $t$ a UTC week, and $m$ an
outcome component. Two objects are kept apart. The \emph{economic object} is LP welfare,
the mixture of Eq.~\eqref{eq:welfare}, $W_{it} = \mathrm{FeeRev}^{\mathrm{arb}}_{it}
+ \mathrm{FeeRev}^{\mathrm{fund}}_{it} - \mathrm{LVR}_{it}$; it motivates the question but
is not directly identified, since fee-revenue attribution, LVR, trader types, and the
arbitrage/fundamental split are partly latent. The \emph{statistical object} is the vector of reconstructed outcomes
$Y_{it} = (Y^m_{it} : m \in \mathcal{M})$ defined in Section~\ref{sec:data}: primary LP
supply $\mathcal{M}_A$, composition and lifecycle $\mathcal{M}_B$, and the feedback chain
$\mathcal{M}_C$. The paper estimates effects on $Y^m$ and uses them to bound and diagnose
$W$, which it does not directly identify.

The fee variables and the assignment fix the notation. The posted trader-facing fee tier
$c_i$ prices execution; the LP take-rate $\rho_{it} \in (0,1]$ is the LP share of collected
fees; the protocol-fee switch sets $\rho_i = 1 - \kappa_i$ with tier-assigned intensity
$\kappa_i \in \{1/4, 1/6\}$ and leaves $c_i$ fixed. Write $D_i$ for treatment, $t_{0i}$ for
the pool-specific activation week, $X_{i0}$ for pre-treatment covariates, $S_i$ for the
pre-period realized fee revenue at the coverage list margin, and $U_i$ for latent pool
fundamentals. The switch gives support for the LP-side take-rate channel but not for
trader-facing dynamic-fee effects, whose intervention $c_i$ does not vary and whose trader
and routing variables are latent.

For each component $m$, let $Y^m_{it}(\bar{\rho}_{i,1:t})$ denote the potential outcome at
week $t$ under the take-rate path $\bar{\rho}_{i,1:t}$ up to that week, allowing carryover
and adjustment costs. Write the observed \emph{switch} path
$\bar{\rho}^{\mathrm{sw}}_{i,1:t}$ ($\rho_i = 1 - \kappa_i$ after activation) against the
counterfactual \emph{no-switch} path $\bar{\rho}^{\mathrm{no}}_{i,1:t}$ ($\rho_i = 1$
throughout). The primary estimand is the finite-horizon dynamic matched-overlap average
treatment effect on the treated (ATT) for component $m$,
\begin{equation}
\Delta^m_h \;=\; \E\bigl[\, Y^m_{i, t_{0i}+h}(\bar{\rho}^{\mathrm{sw}}_{i,1:t_{0i}+h})
- Y^m_{i, t_{0i}+h}(\bar{\rho}^{\mathrm{no}}_{i,1:t_{0i}+h})
\,\bigm|\, D_i = 1,\ i \in \mathcal{M} \,\bigr],
\label{eq:att}
\end{equation}
where $\mathcal{M}$ is the matched-overlap subpopulation of Section~\ref{sec:designsec}.
The main paper focuses on $m \in \mathcal{M}_A$, especially active liquidity and depth.
$\Delta^m_h$ is finite-horizon and reduced-form: it includes carryover, adjustment costs,
and endogenous feedback up to horizon $h$, and is neither a static one-period effect nor
an infinite-horizon steady-state or total welfare effect. $K_L$ is shorthand for this
reduced-form LP-supply response on $\mathcal{M}_A$, not a structural partial-equilibrium
elasticity holding demand fixed; because the $\mathcal{M}_C$ margins move endogenously
with the shock, they are reported as outcomes, not controlled away. Tier-assigned intensity
variation ($\kappa_i \in \{1/4, 1/6\}$) is retained only as a secondary diagnostic and is
not interpreted as a structural tax elasticity.

\subsection{Matched-overlap event study}\label{sec:method}\label{sec:designsec}

For each component $m$, the primary specification is the event study, following the
matched-panel difference-in-differences and event-study literature
\citep{roth2023did,sunabraham2021}.
\begin{equation}
Y^m_{it} = \alpha^m_i + \delta^m_t + \sum_{k \neq -1} \beta^m_k\,
\mathbf{1}[t - t_{0i} = k]\, D_i + \sum_{\tau} \mathbf{1}[t = \tau]\,
X_{i0}'\theta^m_\tau + \varepsilon^m_{it},
\label{eq:es}
\end{equation}
with pool effects $\alpha^m_i$, week effects $\delta^m_t$, pool-specific event time
$t_{0i}$, and the reference bin $k = -1$ omitted, so all $\beta^m_k$ are relative to the
last clean pre-activation week. $X_{i0}$ collects \emph{pre-treatment} baselines only:
pre-period realized fee revenue $S_i$, volume, volatility, pool age, pair class, and tier.
These enter through matching and calendar-time interactions, never as additive controls
the pool effects would absorb; the proposal-to-activation weeks carry their own event-time
bins and are never pooled into the clean pre-period. The sequence $\{\beta^m_k\}$ is the
reduced-form event-study response path: IRF-like (a horizon-by-horizon path after a
discrete policy shock, in the spirit of local projections \citep{jorda2005}), not a
structural VAR impulse response and not an infinite-horizon equilibrium effect. For
$m \in \mathcal{M}_A$ this is the main response path; for $m \in \mathcal{M}_B,
\mathcal{M}_C$ it is the mechanism path, reported as such and not as causal mediation.
Post-treatment volume, fee revenue, slippage, depth, JIT, and LP composition are outcomes,
never main-specification controls. The pool effect absorbs everything constant within a
pool (token pair, tier, the time-invariant $c_i$); the week effect absorbs market-wide
shocks.

The matched-overlap sample $\mathcal{M}$ is built on pre-period data only
\citep{rosenbaum1983,stuart2010}: exact matching
on fee tier and pair class; nearest neighbours ($k \leq 3$, caliper 0.5 log-points) on the
selection variable itself, pre-period realized fee revenue $S_i$; controls restricted to
low spillover exposure; treated units without overlap excluded from $\mathcal{M}$ and
reported. Exposure is made explicit through a pre-treatment matrix $A_{ij}$ (pair overlap
$\times$ router substitutability); a control's aggregate exposure is
$E_i = \sum_j D_j A_{ij}$, and primary controls must be low-exposure. Controls with material
exposure are excluded from the matched counterfactual rather than treated as clean controls. The scope is finite-horizon
and reduced-form \citep{bojinov2019}: the design estimates the response paths
$\{\Delta^m_h\}$ up to the reported horizons and does not identify the infinite-horizon
steady-state or a total welfare decomposition, which would require a structural dynamic
model the paper does not claim.

\subsection{Identification and inference}\label{sec:idmap}\label{sec:assumptions}

The dynamic estimand $\Delta^m_h$ becomes the event-study coefficient of
Eq.~\eqref{eq:es} under dynamic parallel trends \emph{in untreated increments}. Along
the no-switch path, for matched treated pools and matched low-exposure controls and
conditional on $X_{i0}$, when treated and control increments track after the
baseline-time adjustment, the estimand equals the observable difference-in-differences
contrast
\begin{equation}
\Delta^m_h = \E\!\bigl[Y^m_{i,t_{0i}+h}-Y^m_{i,t_{0i}-1}\mid D_i{=}1, i\in\mathcal{M}\bigr]
- \E\!\bigl[Y^m_{i,t_{0i}+h}-Y^m_{i,t_{0i}-1}\mid D_i{=}0, i\in\mathcal{M}\bigr].
\label{eq:did}
\end{equation}
Eq.~\eqref{eq:es} is the sample projection analog of this contrast, with the single
reference period replaced by the full clean pre-period and with pool, week, matching, and
baseline-time structure absorbing levels and calendar. The coefficient $\beta^m_h$
therefore targets $\Delta^m_h$ only when the assumptions below hold; otherwise the event
path is a measurement object without a causal reading.

The framework is an observational panel of potential outcomes, not a randomized
experiment. Writing the untreated-outcome projection
\begin{equation}
Y^m_{it}(\bar\rho^{\mathrm{no}}) = \alpha^m_i + \delta^m_t + X_{i0}'\theta^m_t
+ \varepsilon^m_{it}(0),
\label{eq:proj}
\end{equation}
the core restriction is conditional mean parallel trends in these untreated increments,
\emph{not} full unconfoundedness:
\begin{equation}
\begin{aligned}
&\E\!\bigl[\varepsilon^m_{i,t_{0i}+h}(0)-\varepsilon^m_{i,t_{0i}-1}(0)
\mid D_i{=}1, i\in\mathcal{M}, X_{i0}\bigr] \\
&\qquad= \E\!\bigl[\varepsilon^m_{i,t_{0i}+h}(0)-\varepsilon^m_{i,t_{0i}-1}(0)
\mid D_i{=}0, i\in\mathcal{M}, X_{i0}\bigr],
\end{aligned}
\label{eq:ptres}
\end{equation}
after the exposure restrictions and baseline-time adjustment. Inference is
sampling/model-based: $\varepsilon^m_{it}(0)$ may be serially correlated within a pool and
cross-sectionally correlated within a token-pair / route-substitution cluster, so inference
clusters at the token-pair level with a restricted wild cluster bootstrap
\citep{liang1986,cameronmiller2015,cameron2008wild,roodman2019boottest}. Randomization
inference over placebo activation dates is a timing-sensitivity diagnostic, not a claim of
literal random assignment.

The assumptions under which $\beta^m_h$ carries a causal reading are compact. Let
$e_{i,1:t}$ be the exposure path of pool $i$; the primary design restricts controls to low
exposure and routes exposed units to the spillover design.

\noindent\textbf{A1} (replayability): $D_i$, $\kappa_i$, $t_{0i}$, unit role, and each
$Y^m_{it}$ are deterministic functions of raw logs and fixed reconstruction rules.

\noindent\textbf{A2} (matched overlap): a treated unit enters $\mathcal{M}$ only if a
low-exposure control exists within the pre-specified caliper on $(S_i, X_{i0})$.

\noindent\textbf{A3} (dynamic conditional parallel trends): the no-switch increments of
Eq.~\eqref{eq:ptres} track for treated and low-exposure controls after baseline-time
adjustment.

\noindent\textbf{A4} (restricted interference): primary controls satisfy $e_i \approx 0$
\citep{aronow2017}; exposed units route to the spillover design, whose sign is not imposed
ex ante, so the primary estimate is interpreted for the low-exposure matched-overlap
population.

\noindent\textbf{A5} (no clean-pre anticipation): the clean pre-period excludes the
proposal-to-activation weeks; if future policy information predicts pre-proposal
treated--control divergence, the causal reading fails.

\noindent Under A1--A5, $\beta^m_h$ is the adjusted event-study analog of $\Delta^m_h$; if
an assumption fails, the event path is descriptive. Credibility is assessed through
reconstruction checks (Appendix~\ref{app:impl}), the parallel-trend and separated-anticipation
gates, placebo dates, caliper and window robustness, Honest-DiD bounds \citep{rambachan2023}
and an omitted-confounding robustness value \citep{cinelli2020}, and reduced-form mechanism
paths reported as outcomes rather than mediation. These diagnostics do not select
specifications \citep{roth2022pretest}; they determine whether an outcome receives a causal
or a descriptive label. Claims are routed by observability and support: the LP-side
take-rate response is design-based, while trader-facing dynamic-fee protection is not
identified and is routed to a model-conditioned diagnostic or a non-estimand.

\section{Results}\label{sec:results}

\begin{figure}[H]\centering
\includegraphics[width=0.92\textwidth]{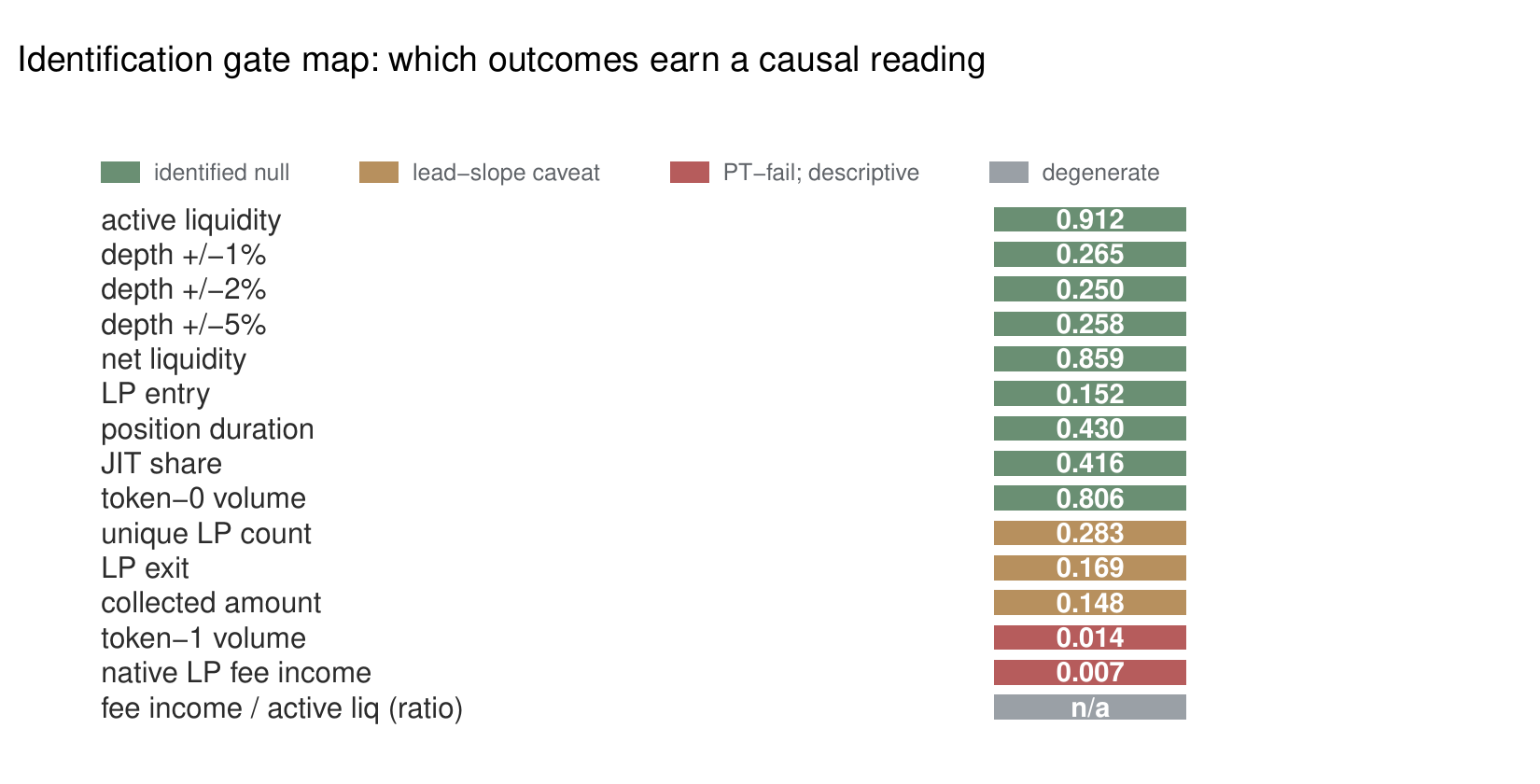}
\caption{Identification gate map for the outcome family. Tile colour is the causal
disposition and the number is the joint pre-trend $p$. Liquidity, depth, and participation
outcomes earn an identified-null reading; token-1 volume and native fee income fail the
pre-trend gate and stay descriptive; the fee-income-per-active-liquidity ratio is
degenerate. Lead-slope and post-CI detail are in Appendix~\ref{app:smd}.}
\label{fig:gatemap}
\end{figure}

No post-period estimate is reported before the frozen panel is built, the completeness and
sanity checks logged, and the pre-trends reviewed. A reading that survives the credibility
framework of Section~\ref{sec:assumptions} earns a causal interpretation; one that does not
is reported descriptively. Figure~\ref{fig:gatemap} maps the outcome family to its
identification disposition; the remainder of this section reports the identified results.

\subsection{Main LP-supply and depth response}\label{sec:res-main}

\begin{figure}[t]\centering
\begin{subfigure}[t]{0.48\textwidth}\centering
\includegraphics[width=\textwidth]{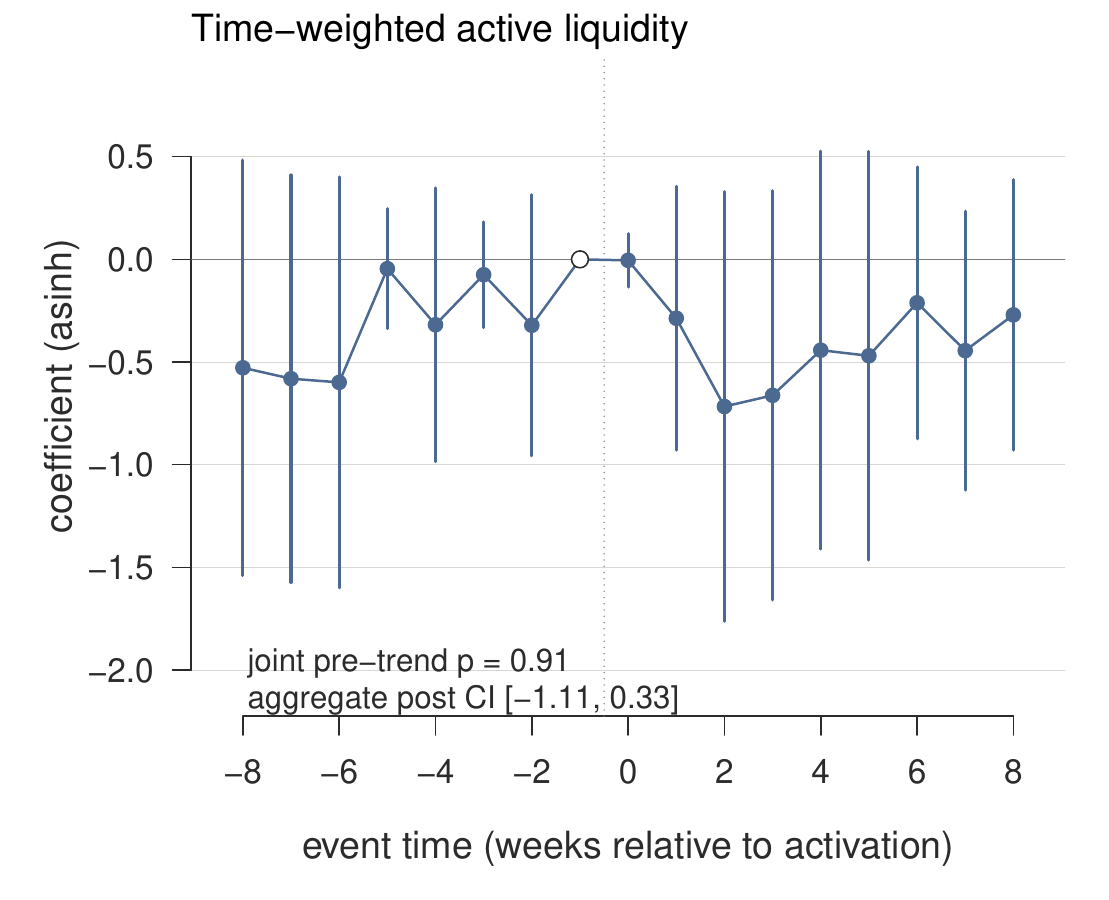}
\subcaption{Time-weighted active liquidity.}\label{fig:mainpaths-a}
\end{subfigure}\hfill
\begin{subfigure}[t]{0.48\textwidth}\centering
\includegraphics[width=\textwidth]{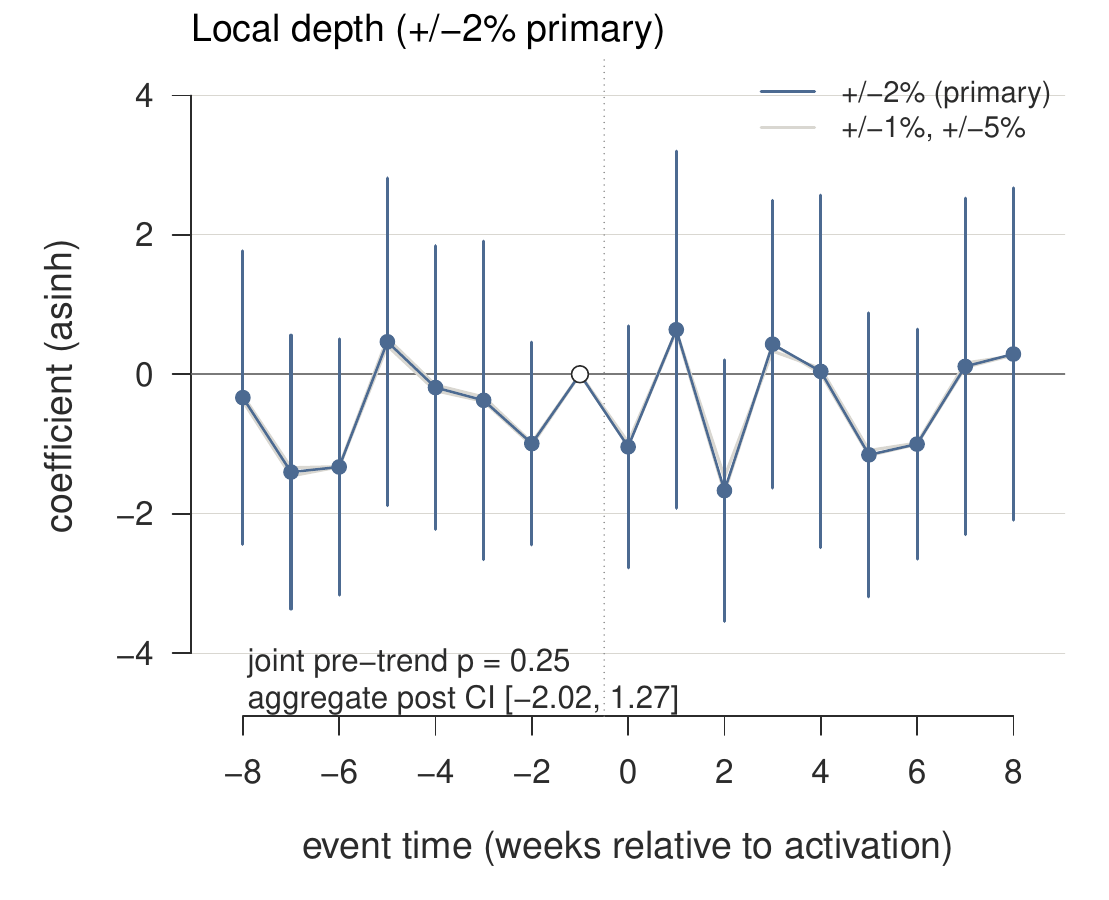}
\subcaption{Local depth ($\pm 2\%$ primary).}\label{fig:mainpaths-b}
\end{subfigure}
\caption{Main event-study paths, primary $\pm 8$-week window, reference $k=-1$, on the
inverse-hyperbolic-sine scale, with token-pair CR1 $95\%$ intervals. (a) Time-weighted
active liquidity. (b) Local depth ($\pm 2\%$ primary, with $\pm 1\%$ and $\pm 5\%$ as
thin grey paths). Intervals include zero throughout the post period: a non-detection at the
design's resolution, not a precise zero. Disposition: identified null.}
\label{fig:mainpaths}
\end{figure}

The matched-overlap design clears the
identification gate for the LP-supply and liquidity-quality outcomes. In the primary
$\pm 8$-week window the estimation sample contains 779 matched treated pools, 303 matched
controls, 17{,}598 pool-weeks, and 1{,}013 token-pair clusters. Joint lead tests do not
reject parallel pre-trends for the primary liquidity object, time-weighted active
liquidity and depth at the $\pm 1\%$, $\pm 2\%$, and $\pm 5\%$ bands, nor for the
secondary LP-composition family (net liquidity, LP entry and exit, unique-LP count,
position duration, same-block JIT share, and collected amount). The primary outcomes are
clean on the secondary lead-slope diagnostic; among the composition outcomes a few carry
mild slope warnings, most consistently collected amount and at the margin unique-LP
count, but their joint pre-trend gates pass and their post-period intervals include zero.
Across the identified family the aggregate post-period intervals of
Eq.~\eqref{eq:es} include zero on the inverse-hyperbolic-sine scale: for
time-weighted active liquidity the interval is $[-1.11, 0.33]$, and for the central depth
band ($\pm 2\%$) it is $[-2.02, 1.27]$. The design does not detect an average short-run
withdrawal of liquidity, loss of local depth, or change in LP composition among
matched-overlap treated pools after the switch (Figure~\ref{fig:mainpaths}).

\subsection{Precision and robustness}\label{sec:res-precision}

\begin{figure}[t]\centering
\includegraphics[width=0.80\textwidth]{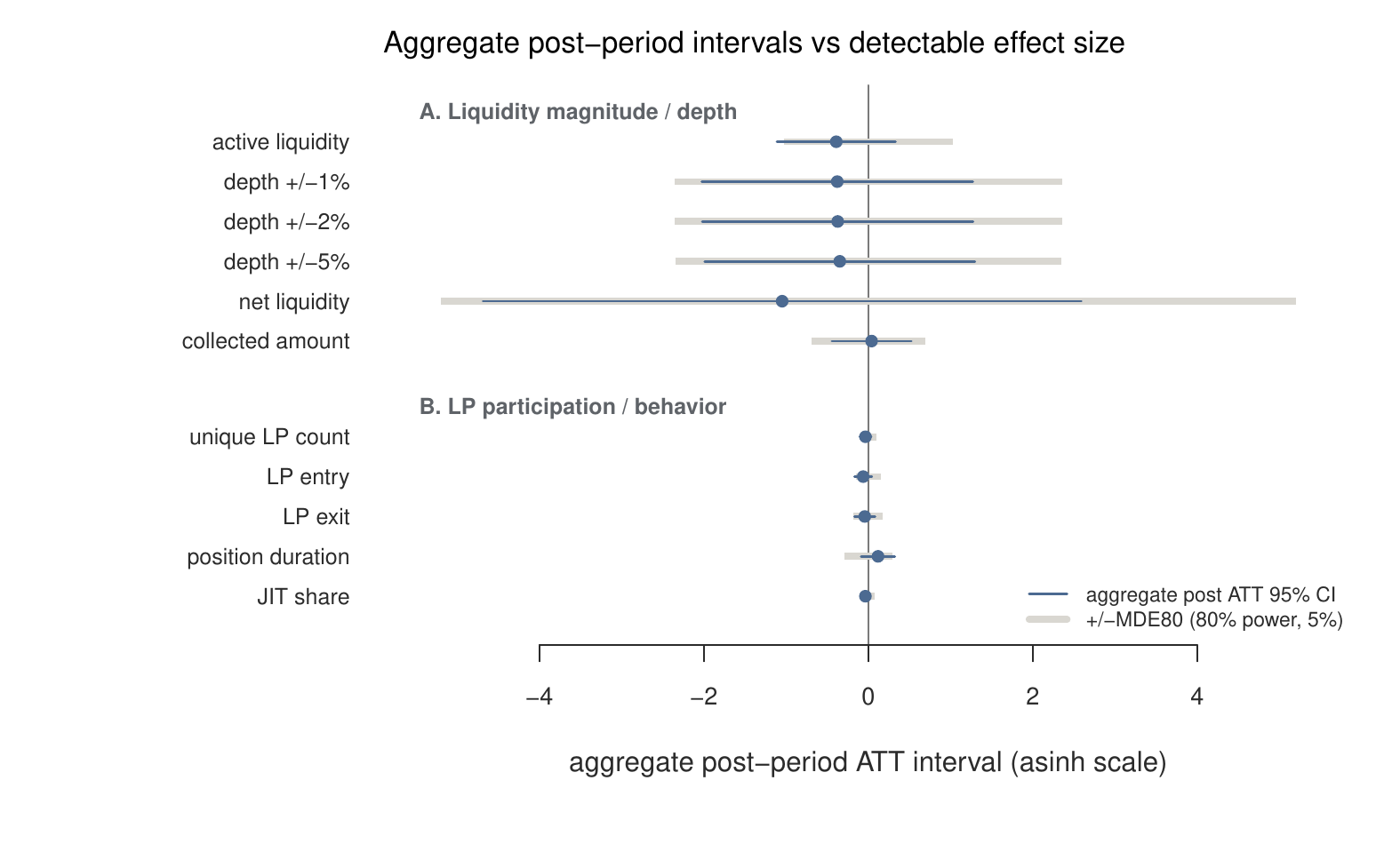}
\caption{Aggregate post-period ATT intervals (inverse-hyperbolic-sine scale) against the
$\pm$MDE$_{80}$ band (80\% power, 5\%) for the identified outcomes, grouped into (A)
liquidity magnitude / depth and (B) LP participation / behavior. Participation and behavior
are sharply estimated; liquidity magnitude and depth are less precise. Disposition:
identified null (precision).}
\label{fig:precision}
\end{figure}

This is a non-detection at the design's resolution, not
a high-precision zero. On the inverse-hyperbolic-sine scale (approximately a log scale at
these magnitudes; see \citealp{burbidge1988,bellemare2020asinh}) the active-liquidity interval is a multiplicative range of about
$[0.33, 1.39]$, with an $80\%$ minimum detectable effect (MDE) near $1.03$; the depth
interval is wider, about $[0.13, 3.58]$ with an MDE near $2.35$. The design thus rules out
only very large average responses for active liquidity, and depth is less precisely
estimated. LP participation and behavior are the sharpest evidence: the entry, exit,
unique-LP, and JIT-share nulls are tightly estimated (MDEs near $0.1$ asinh units,
multiplicative ranges within about $\pm 10\%$), so the absence of a participation or
toxic-flow response is well powered, whereas net liquidity is estimated too imprecisely to
be informative (Figure~\ref{fig:precision}). We therefore report a well-powered no-change
statement for LP participation, and, for liquidity magnitude and depth, the absence of a
large average response over this short horizon without excluding moderate ones.

The null survives the pre-specified sensitivity set. Honest-DiD relative-magnitude bounds
have breakdown value zero because the aggregate intervals already include zero, leaving no
significant estimate to overturn. Placebo-date runs are flat for the primary liquidity and
depth outcomes. Entropy balancing \citep{hainmueller2012} removes the residual pre-period
fee-revenue imbalance without changing the sign or significance of the primary estimates.
The conclusion is stable to the event window ($\pm 6$, $\pm 8$, $\pm 12$ weeks) and to the
matched-overlap caliper: regenerating the design at calipers $0.25$, $0.5$, and $1.0$ shifts
the support set as expected, but all eleven identified outcomes continue to pass the joint
pre-trend gate and return intervals containing zero. The per-caliper table and MDE column
are in Appendix~\ref{app:smd}.

\FloatBarrier

\subsection{Flow and revenue outcomes fail the causal gate}\label{sec:res-flow}

The flow and revenue outcomes do not receive the
same causal reading. Token-1 volume and native LP fee income fail the primary
parallel-trends gate, and allowing matched-stratum-by-week trends does not repair the
failure. This is consistent with the treatment list, since these outcomes sit closest to the
pre-period fee-revenue selection margin and the matched controls do not supply a credible
untreated trend for them. They are reported descriptively; the per-active-liquidity
fee-income ratio is degenerate in the reconstructed panel and is not informative. A
pre-registered auxiliary factor-model check for token-1 volume also failed its validation
gate, so the flow and fee-income outcomes remain descriptive; the diagnostic detail is in
the repository.

\section{Conclusion}\label{sec:conclusion}

Under the frozen matched-overlap design, the protocol-fee switch that cut LP take-rates
produces no large short-run average response in active liquidity or local depth; LP
participation and composition are more precisely estimated and also show no detectable
response. The result is a non-detection at the design's resolution rather than a precise
zero, and it survives Honest-DiD, placebo-date, entropy-balancing, event-window, and
caliper sensitivity. Token-1 volume and native fee income fail the identification gate and
are reported descriptively; a pre-registered auxiliary factor-model counterfactual for
volume did not pass its validation gate and is not given a causal reading.

The same record also marks the boundary of the claim. Because the trader-facing fee $c_i$
does not vary, and trader types and router decision sets are latent, the switch does not
identify trader-facing dynamic-fee protection. LP-side supply and depth outcomes receive the
design-based reading (Figure~\ref{fig:gatemap}); flow and fee-income outcomes remain
descriptive after failing the gate. Not every DeFi fee claim is identifiable from on-chain
history; the audit estimates the one that is and labels the rest as boundary, diagnostic, or
bound.

For fee-controller evaluation, the implication is narrow: simulations that freeze liquidity
supply assume away $K_L$. The estimates here constrain the short-run LP-exit margin, but do
not identify total welfare, long-run elasticity, or trader-side dynamic-fee protection.

\FloatBarrier
\bibliography{references}

\appendix

\section{Match balance and sensitivity}\label{app:smd}\label{app:sens}

\begin{figure}[H]\centering
\begin{subfigure}[t]{0.48\textwidth}\centering
\includegraphics[width=\textwidth]{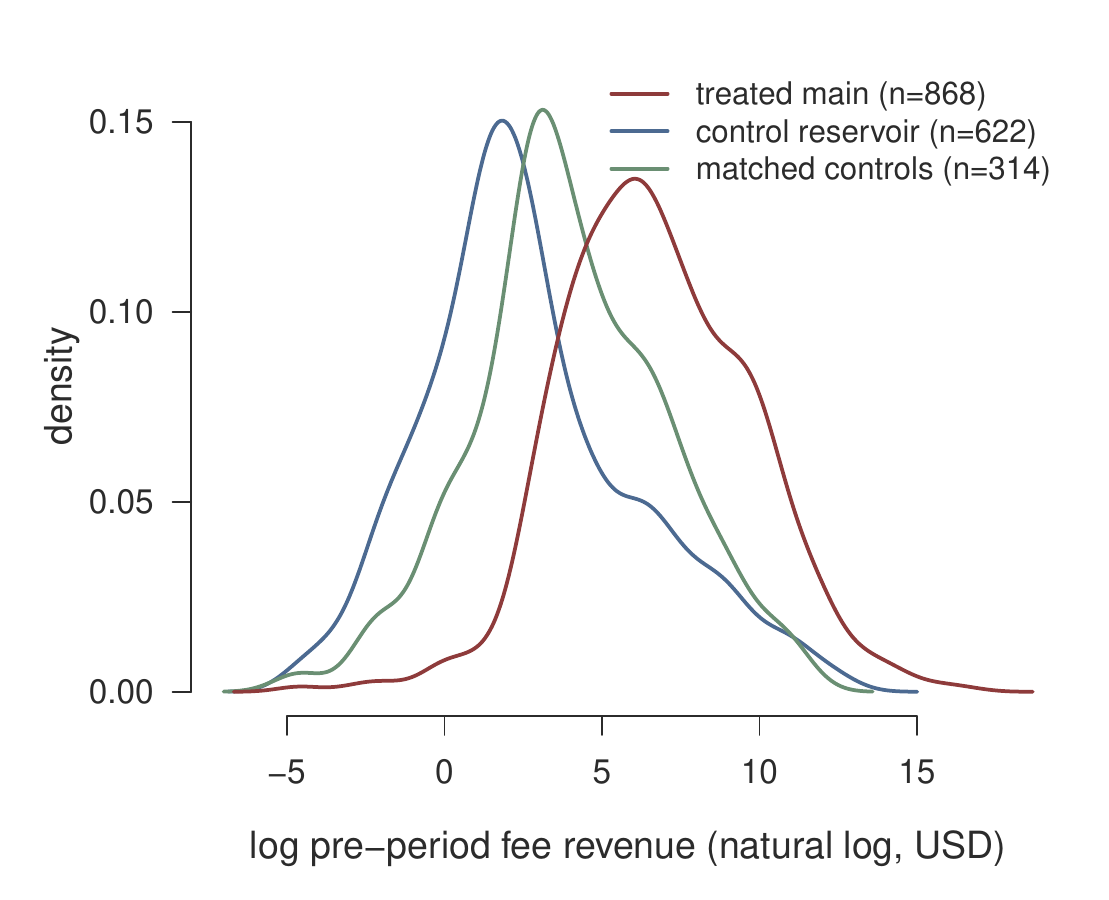}
\subcaption{Fee-revenue overlap.}\label{fig:balance-a}
\end{subfigure}\hfill
\begin{subfigure}[t]{0.48\textwidth}\centering
\includegraphics[width=\textwidth]{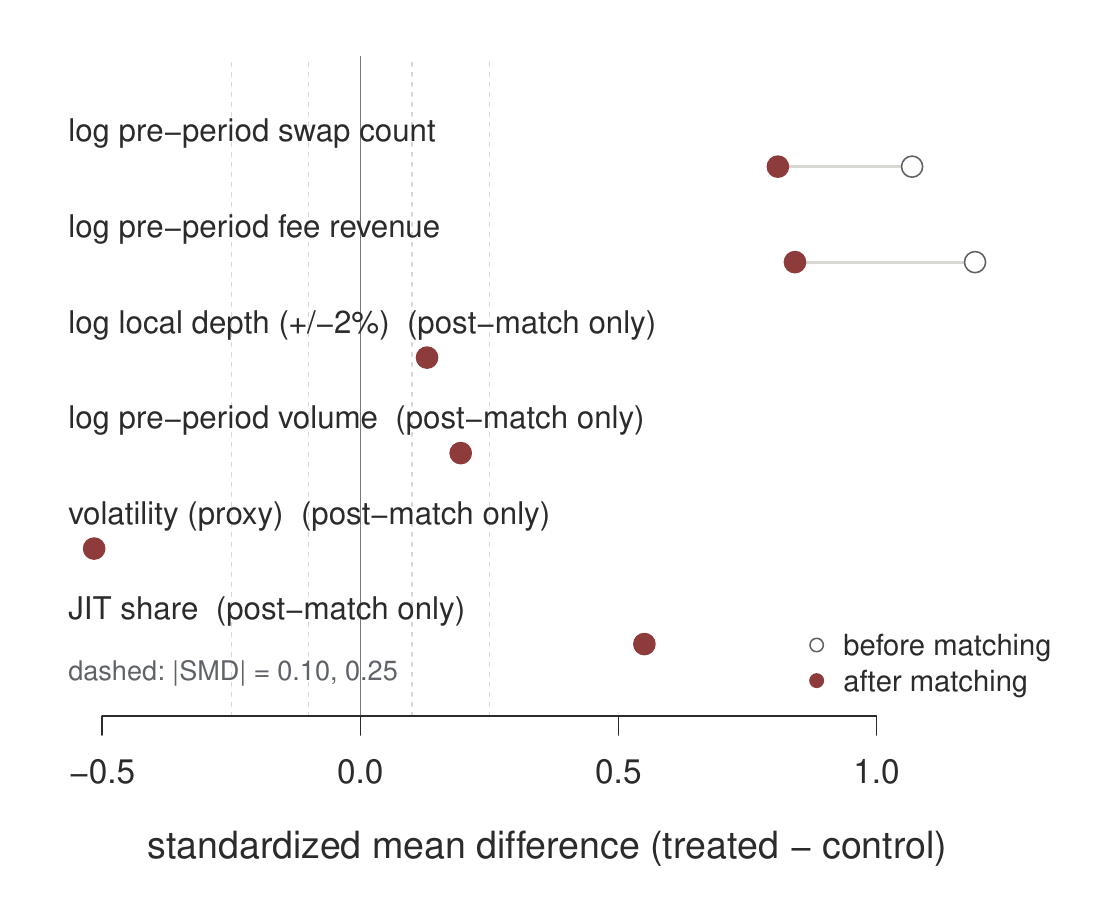}
\subcaption{Covariate love plot.}\label{fig:balance-b}
\end{subfigure}
\caption{Matched-overlap balance and support. (a) Distribution of log pre-period
realized fee revenue for treated main, control reservoir, and matched controls. (b)
Signed standardized-mean-difference love plot; the fee-revenue and swap-count matching
covariates carry before/after markers, while the remaining covariates show post-match
balance only (their pre-match reservoir trajectories lie outside the frozen panel).
Residual imbalance remains on the selection margin; identification rests on the
event-study gates, not level balance. Disposition: design support.}
\label{fig:balance}
\end{figure}

The matched-overlap design achieves support but not level balance: matching narrows the
$\log$ pre-period fee-revenue gap from a standardized mean difference of $1.19$ to $0.84$
(swap count $1.07$ to $0.81$), leaving residual imbalance on the coverage margin, so
identification rests on conditional parallel trends with baseline-time interactions rather
than on level balance (Figure~\ref{fig:balance}).

Table~\ref{tab:caliper} reports the identified LP-side family across the three
matched-overlap calipers at the primary $\pm 8$-week window. Every outcome passes the joint
pre-trend gate and returns an interval containing zero at all three calipers; the primary
liquidity/depth intervals are wide (the null is a non-detection, not a precise zero), while
the participation and JIT-share intervals are tight.

\begin{table}[H]\centering\small
\caption{Identified LP-side family: pre-trend and null across matched-overlap calipers,
primary $\pm 8$-week window. Disposition: identified null (robustness).}\label{tab:caliper}
\begin{tabular}{lcccl}
\toprule
Outcome & pre-trend $p$ (0.25/0.5/1.0) & ATT CI (asinh) & MDE$_{80}$ & disposition \\
\midrule
active liquidity (TWL) & 0.90/0.91/0.82 & $[-1.11,0.33]$ & 1.03 & primary \\
depth $\pm1\%$ & 0.25/0.26/0.25 & $[-2.03,1.27]$ & 2.36 & primary \\
depth $\pm2\%$ & 0.23/0.25/0.24 & $[-2.02,1.27]$ & 2.35 & primary \\
depth $\pm5\%$ & 0.24/0.26/0.23 & $[-1.99,1.30]$ & 2.35 & primary \\
net liquidity & 0.53/0.86/0.82 & $[-4.69,2.59]$ & 5.20 & secondary \\
unique LP count & 0.26/0.28/0.43 & $[-0.10,0.03]$ & 0.10 & secondary$^{\dagger}$ \\
LP entry & 0.19/0.15/0.28 & $[-0.17,0.04]$ & 0.15 & secondary \\
LP exit & 0.14/0.17/0.23 & $[-0.17,0.08]$ & 0.18 & secondary \\
position duration & 0.52/0.43/0.38 & $[-0.09,0.32]$ & 0.29 & secondary \\
JIT share & 0.48/0.42/0.35 & $[-0.09,0.02]$ & 0.07 & secondary \\
collected amount & 0.30/0.15/0.26 & $[-0.44,0.52]$ & 0.69 & secondary$^{\dagger}$ \\
\bottomrule
\end{tabular}
\end{table}
{\footnotesize $^{\dagger}$ carries a mild lead-slope caveat (the joint pre-trend gate
passes and the interval contains zero). Design match counts by caliper:
$0.25$ (696 matched treated / 296 controls), $0.5$ (786 / 314), $1.0$ (825 / 323); the
$\pm 8$ estimation sample is smaller after windowing and singleton fixed-effect removal
(at $0.5$: 779 treated / 303 controls, 17{,}598 pool-weeks, 1{,}013 clusters). Honest-DiD,
placebo-date, and entropy-balancing sensitivity (Section~\ref{sec:res-precision}) leave the
identified-null dispositions unchanged.}

\section{Reconstruction and reproducibility}\label{app:impl}

\begin{figure}[H]\centering
\includegraphics[width=0.92\textwidth]{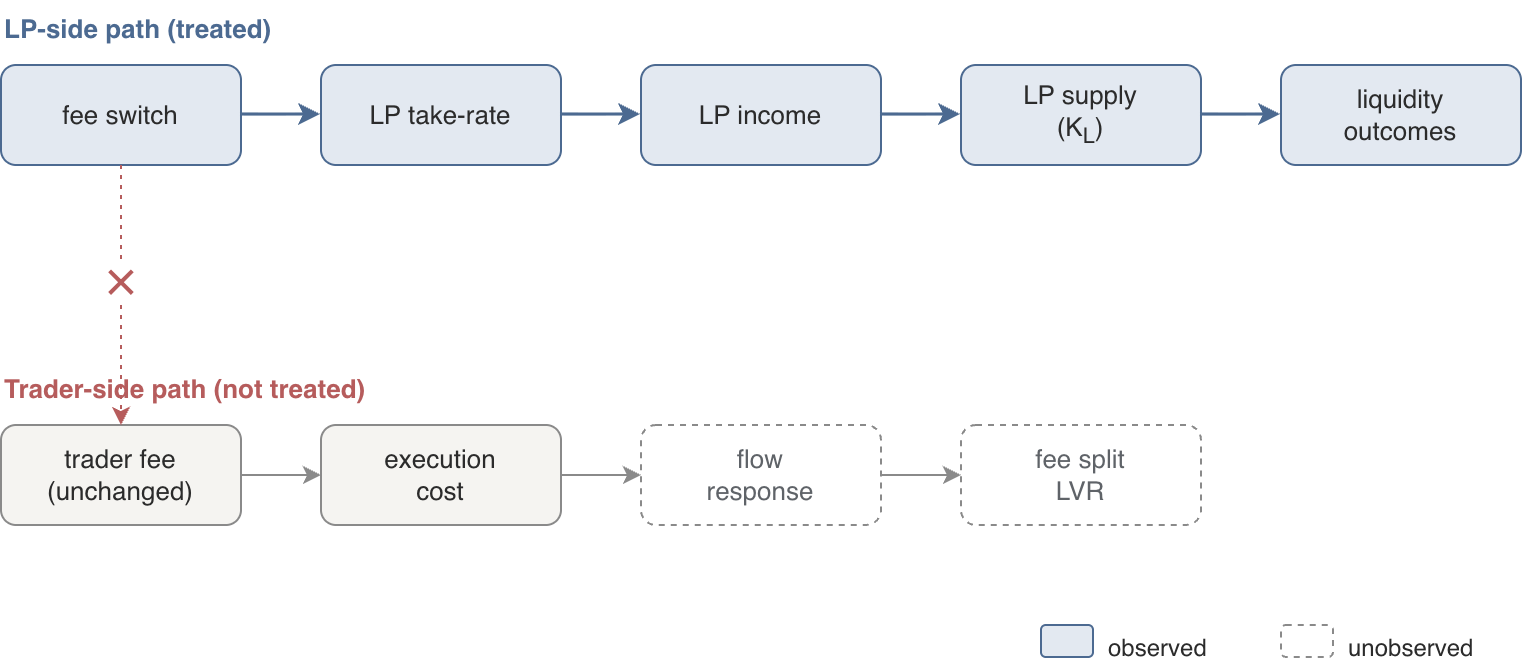}
\caption{The channel map. Solid nodes are measured or reconstructed; the LP-side lane is
activated by the switch ($\rho_{it} \to 1 - \kappa_i$) and enters the design-based analysis;
dashed nodes require latent trader or routing information and are not primary evidence; the
crossed edge marks the missing treatment variation ($c_i$ unchanged). Observability sorts
each variable into measured, reconstructed, or latent; support records whether its
intervention varies and at what level.}
\label{fig:channels}
\end{figure}

Figure~\ref{fig:channels} gives the channel map that sorts each variable by observability and
support. Treatment, event time, unit roles, and outcomes are deterministic functions of raw
logs; intermediate panels found invalid on inspection were quarantined and used for neither
pre-trend diagnostics nor any estimate.

The frozen outcome construction proceeds from raw normalized Swap/Mint/Burn/Collect events
through tick-book initialization (so depth is not measured against an empty book), then
time-weighted active liquidity split at UTC week boundaries, depth within $\pm 1\%$,
$\pm 2\%$, and $\pm 5\%$ bands, the LP lifecycle by owner-range key, same-block JIT share,
and LP fee income scaled by per-token decimals, with an artifact manifest carrying a SHA256
hash per file.

A self-contained Rust toolchain performs this reconstruction, memory-bounded for the full
seventy-million-event history; its output is checked against an independent reconstruction,
row-for-row and exact, on a diverse pool subset (subset bit-parity), and the Rust
event-study point estimates are checked against the R inference (estimator parity). Inference
uses established R packages (\texttt{fixest}, \citealp{berge2026fixest};
\texttt{fwildclusterboot}, \citealp{fischer2021fwildclusterboot}; \texttt{HonestDiD},
\citealp{honestdid_pkg}; \texttt{sensemakr}, \citealp{sensemakr_pkg}; \texttt{ebal},
\citealp{ebal_pkg}; \texttt{fect}, \citealp{liu2024fect,fect_pkg}; \texttt{fdapace},
\citealp{fdapace_pkg}), each estimate recorded with its sample, transform, window, cluster,
seed, and gate outcome. The full reconstruction and inference code is available at
\url{https://github.com/egpivo/amm-lab}.

\FloatBarrier

\end{document}